\documentclass[10pt, conference, compsocconf]{IEEEtran}
\IEEEoverridecommandlockouts
\usepackage{flushend}
\usepackage{caption}
\usepackage{subfigure}
\usepackage{subcaption}
\usepackage{graphicx}
\usepackage[table, svgnames, dvipsnames]{xcolor}
\usepackage{booktabs}
\usepackage{comment}
\usepackage{cite}
\usepackage[utf8]{inputenc}
\usepackage{amsmath,amssymb,amsfonts}
\usepackage{soul}
\definecolor{green}{rgb}{0.16, 0.67, 0.53}
\usepackage{textcomp,balance,url}
\usepackage[left=1.62cm,right=1.62cm,top=1.9cm]{geometry}
\definecolor{chestnut}{rgb}{0.97, 0.51, 0.47}
\def\BibTeX{{\rm B\kern-.05em{\sc i\kern-.025em b}\kern-.08em
    T\kern-.1667em\lower.7ex\hbox{E}\kern-.125emX}}

\definecolor{blue}{rgb}{0.0, 0.47, 0.75}
\usepackage{tikz}
\usetikzlibrary{positioning, arrows.meta}

\DeclareMathOperator*{\argmin}{\arg\min}

\begin{document}
\title{Resilient by Design -- Active Inference for Distributed Continuum Intelligence}
\author{
\IEEEauthorblockN{Praveen Kumar Donta\IEEEauthorrefmark{1}, Alfreds Lapkovskis\IEEEauthorrefmark{1}, Enzo Mingozzi\IEEEauthorrefmark{2}, and Schahram Dustdar\IEEEauthorrefmark{3}}
\IEEEauthorblockA{\IEEEauthorrefmark{1}\textit{Department of Computer Systems and Sciences}, \textit{Stockholm University,} Stockholm 164 25, Sweden.} 
\IEEEauthorblockA{\IEEEauthorrefmark{2}\textit{Department of Information Engineering}, \textit{University of Pisa,} Pisa, Italy.} 
\IEEEauthorblockA{\IEEEauthorrefmark{3}\textit{Distributed Systems Group, TU Wien}, Vienna 1040, Austria and \textit{ICREA} Barcelona 08002, Spain}
 \texttt{\{praveen,alfreds.lapkovskis\}@dsv.su.se}, \texttt{enzo.mingozzi@unipi.it}, \texttt{dustdar@dsg.tuwien.ac.at}
}

\maketitle

\begin{abstract}
Failures are the norm in highly complex and heterogeneous devices spanning the distributed computing continuum (DCC), from resource-constrained IoT and edge nodes to high-performance computing systems. Ensuring reliability and global consistency across these layers remains a significant challenge, particularly for AI-driven workloads that require real-time, adaptive coordination. This work-in-progress paper introduces a Probabilistic Active Inference Resilience Agent (PAIR-Agent) to achieve resilience in DCC systems. PAIR-Agent performs three core operations: (i) constructing a causal fault graph from device logs, (ii) identifying faults while managing certainties and uncertainties using Markov blankets and the free energy principle, and (iii) autonomously healing issues through active inference. Through continuous monitoring and adaptive reconfiguration, the agent maintains service continuity and stability under diverse failure conditions. Theoretical validations confirm the reliability and effectiveness of the proposed framework.
\end{abstract}

\section{Introduction}\label{sec:Introduction}
\IEEEPARstart{N}{o} one can deny that computing infrastructure has transformed dramatically with Generative AI, driving unprecedented demand for high-performance resources spanning sensor nodes, edge devices, network elements, and server clusters. In response, the Distributed Computing Continuum (DCC) paradigm has emerged, orchestrating complex tasks across the full device continuum, allocating smaller, latency-sensitive tasks locally to reduce energy use and preserve privacy, while offloading resource-intensive tasks to high-performance servers \cite{dustdar2022distributed,donta2024human_based}. Such dynamic task distribution optimizes resource use, balances energy demand, and supports scalable operations. As the number and diversity of devices increase, data volumes surge, and workloads become more dynamic, system complexity concurrently grows, intensifying failure risks \cite{donta2023governance}. In general, task processing failures in the DCC occur due to factors such as user-initiated aborts, incorrect inputs, insufficient allocated resources, execution errors, node crashes, or communication failures; In DCC, additional failures include the unpredictability brought by device heterogeneity, mobile and battery-dependent nodes, dynamic network and resource shifts, and large-scale consistency challenges associated with hundreds or thousands of edge devices.

Resilience mechanisms are crucial for modern computing systems, since failures are the norm in these infrastructures. The term resilience is chosen over mere fault tolerance in DCC because these systems must not only tolerate discrete faults but also effectively manage and adapt to a wide range of inherent uncertainties \cite{wang2025perturbation,svobodova2009resilient}. 
In DCC, resilience reflects the system’s capability to sustain continuous operation through proactive adaptation, localized recovery, and graceful degradation across diverse cloud, fog, edge, and IoT layers, where unpredictable workload fluctuations, device mobility, connectivity disruptions, and energy variability are common \cite{sagor2024distressnet,10562327}. To achieve these, this paper designs a novel probabilistic active inference resilience agent (PAIR-Agent), with the following contributions. 
\begin{itemize}
    \item PAIR-Agent builds a probabilistic Causal Fault Graph (CFG) using Bayesian Network Structure Learning (BNSL) through various log data from the devices.
    \item Next, it uses the Markov blanket to identify potential faults from CFGs. Further, PAIR-Agent applies the free energy principle for uncertainty-aware fault inference.
    \item Finally, PAIR-Agent autonomously selects corrective actions, minimizing expected free energy and enabling self-healing through active inference.
\end{itemize}

\section{System Model}\label{sec2}
The DCC ($\mathcal{S}$) are composed of $n$ devices including one or $c$ cloud or data centers $\mathcal{C} =\{C_1, C_2, ..., C_c\}$, a set of $f$ Fog nodes $\mathcal{F} =\{F_1, F_2, ..., F_f\}$, a set of $e$ Edge nodes $\mathcal{E} = \{E_1, E_2, ..., E_e\}$, $m$ mobile devices $\mathcal{M} = \{M_1, M_2, ..., M_m\}$, a set of $\iota$ IoT devices $\Gamma = \{I_1, I_2, ..., I_\iota\}$, and a set of $\varsigma$ sensor nodes (SNs) $\Psi = \{\psi_1, \psi_2, ..., \psi_\varsigma\}$. We assume that $\Psi$ are data producers and they are not capable of computing any data. Therefore, they transmit their data to the nearest base station, i.e., $\Gamma$, $\mathcal{M}$, or $\mathcal{E}$ using wireless communication through Bluetooth Low Energy, Zigbee, Cellular, or Wi-Fi channels \cite{cheikhrouhou2016secure}. 

A set of AI-related computational tasks is defined as $\mathcal{T} = \{T_1, T_2, ..., T_t\}$, 
where each task $T_i = \langle \omega_i, \chi_i, \phi_i, \delta_i \rangle$ denotes workload $\omega_i$, input data dependencies $\chi_i$, a computational mapping or model function $\phi_i : \chi_i \mapsto y_i$, and an execution deadline $\delta_i$. 
The DCC graph $\mathcal{S} = (\mathcal{V}, \mathcal{R})$ consists of nodes $\mathcal{V} = \mathcal{C} \cup \mathcal{F} \cup \mathcal{E} \cup \mathcal{M} \cup \Gamma \cup \Psi$ and communication relationships $\mathcal{R} \subseteq \mathcal{V} \times \mathcal{V}$, where each node $v_j \in \mathcal{V}$ executes a subset of tasks $\mathcal{T}(v_j)$ such that 
$\bigcup_{v_j \in \mathcal{V}} \mathcal{T}(v_j) = \mathcal{T}$. 
Each task execution $T_i(v_j)$ follows a three-phase structure 
$\langle \alpha_i, \beta_i, \gamma_i \rangle$, 
representing task initiation, computation, and completion or checkpointing, respectively. 
In general, these complex AI workflows, such as training, are composed of sequential subtasks, e.g., 
$T_{\text{train}} = T_{\text{data-load}} \triangleright T_{\text{forward}} \triangleright T_{\text{backward}} \triangleright T_{\text{update}}$, 
where each subtask can complete and save progress independently to support partial recovery. Each task establishes a stable checkpoint 
$\zeta_i = \langle s_i, D_i, \theta_i \rangle$, 
recording the intermediate state $s_i$, processed dataset segment $D_i$, and model parameters $\theta_i$, enabling restoration and continuation upon failure.

\section{Resilience in Action}\label{sec3}
\begin{figure*}[t]
    \centering
    \includegraphics[width=1.0\textwidth]{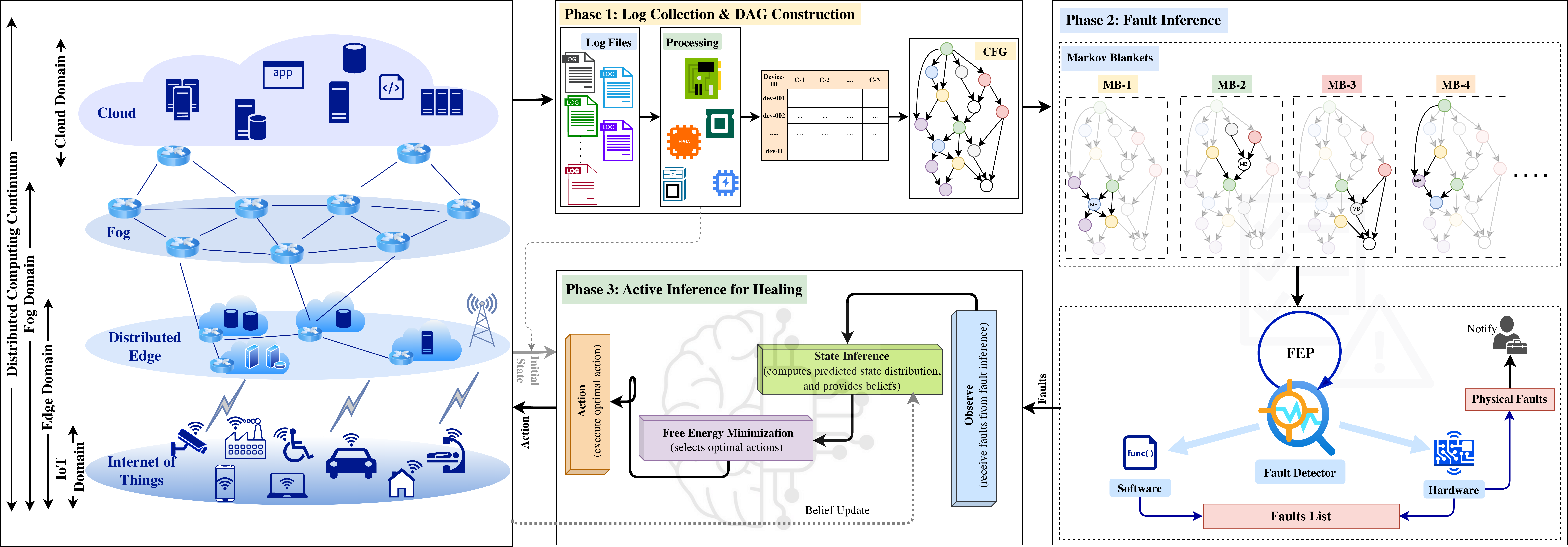}
   \caption{
The proposed PAIR-Agent for resilience in the DCC collects and normalizes logs to form a CFG, infers faults via Markov blanket and FEP analysis, and heals through active inference to restore system stability.
}\label{fig1}
\end{figure*}

The proposed PAIR-Agent iteratively performs the following steps (as depicted in Figure \ref{fig1}): (a) identifying changes in the system logs and collecting data to generate a CFG; (b) performing fault inference by determining certain and uncertain faults using the Markov blanket and free energy principle\cite{kirchhoff2018markov}; and (c) healing the faults through active inference to achieve resilience, as discussed in more detail in the subsequent subsections.

\subsection{Collect and Process Logs}
Initially, the PAIR-Agent coordinates distributed log acquisition across the $\mathcal{S}$ using remote procedure calls (RPCs) to query each device $(v_i \in \mathcal{S})$ for its most recent checkpoint $(\zeta_i^t = \langle s_i^t, D_i^t, \theta_i^t \rangle)$, which encapsulates the local state, processed dataset segment, and learned parameters at analysis round $t$. 
These checkpoints act as temporal anchors that partition system observations into discrete analysis intervals, allowing the agent to process only the log entries generated after $\zeta_i^t$. 
This design ensures synchronization across heterogeneous nodes, prevents redundant ingestion of log segments, tracks round-based progress, and maintains a globally consistent causal timeline of system activity~\cite{saleh2025agentic}. 
Each new round $(t+1)$, therefore, begins with a synchronized collection of incremental log updates $\Delta L_i^{t+1}$, forming a time-evolving dataset that supports continual inference.

After collection, raw logs are parsed and normalized into a structured representation. PAIR-Agent extracts features corresponding to potential fault interpretation or indicators, including user-aborted tasks, resource allocation failures (due to device mobility, computational capacity, or energy level), node crashes, execution-time anomalies, data inconsistencies, and communication disruptions, among others. 
Unlike traditional distributed systems, PAIR-Agent considers each fault as a probabilistic event \(f_i\) embedded within a continuous spatiotemporal context rather than an isolated deterministic failure. 
Each $f_i$ is characterized by a conditional probability distribution $P(f_i \mid \mathbf{x}_i, t)$, where $\mathbf{x}_i$ represents contextual features extracted from the normalized log matrix and $t$ denotes the checkpoint time window. This probabilistic modeling allows the system to capture uncertainty and latent dependencies among hardware, software, and environmental factors influencing fault occurrence across heterogeneous nodes.

To uncover causal dependencies among probabilistic fault variables, PAIR-Agent constructs a time-evolving CFG $\mathcal{G}_{\text{fault}}^{t+1} = (\mathcal{V}_f^{t+1}, \mathcal{E}_f^{t+1})$ using a BNSL. 
Here, each node $v_i \in \mathcal{V}_f^{t+1}$ represents a stochastic variable corresponding to a specific fault or contextual feature extracted from the normalized feature matrix $X^{t+1}$, and each directed edge $e_{ij} \in \mathcal{E}_f^{t+1}$ encodes a conditional dependency $P^{t+1}(v_j \mid \text{Pa}(v_j))$, where $\text{Pa}(v_j)$ denotes the parent set of $v_j$. This structure is inferred incrementally at each checkpoint interval $(t, t+1]$, allowing PAIR-Agent to integrate newly observed evidence $\Delta \mathcal{D}^{t+1}$ while retaining prior knowledge from earlier analysis rounds. 
This is achieved through Hill-Climbing with Bayesian Dirichlet equivalence scoring \cite{scutari2018dirichlet}, combined with probabilistic regularization to handle noise and partial observability in DCC data. 
In scenarios with strong nonlinearity or non-stationary correlations, PAIR-Agent employs variational Bayesian learners to refine causal relations across checkpoints adaptively \cite{friston2003dynamic}.
Each updated Bayesian network $\mathcal{G}_{\text{fault}}^{t+1}$ represents the posterior belief over causal structure given new observations, expressed as $P^{t+1}(\mathcal{G}_{\text{fault}} \mid \mathcal{D}^{t+1}) \propto P(\mathcal{D}^{t+1} \mid \mathcal{G}_{\text{fault}}) \, P^t(\mathcal{G}_{\text{fault}}),$ where $P^t(\mathcal{G}_{\text{fault}})$ encodes the prior structural belief accumulated from previous rounds.

\subsection{Fault Inference}
Building upon the Bayesian CFG $\mathcal{G}_{\text{fault}}^{t+1}$, PAIR-Agent performs probabilistic fault inference by exploiting the local Markov properties of the network. 
For each fault variable $f_i \in \mathcal{V}_f^{t+1}$, the agent defines its Markov blanket $\mathcal{M}(f_i) = \text{Pa}(f_i) \cup \text{Ch}(f_i) \cup \text{Pa}(\text{Ch}(f_i))$, where $\text{Pa}(f_i)$ and $\text{Ch}(f_i)$ denote the parent and child sets of $f_i$, respectively \cite{friston2010free}. 
This blanket represents the minimal sufficient set of variables that render $f_i$ conditionally independent from the rest of the network. 
By monitoring the probabilistic interactions within $\mathcal{M}(f_i)$, the agent computes the posterior probability of fault activation or propagation, i.e. 
\begin{align*}
    P^{t+1}(f_i \mid \mathcal{M}(f_i)) &\propto
    P^{t+1}(f_i \mid \text{Pa}(f_i)) \\ &\times \prod_{c_j \in \text{Ch}(f_i)} P^{t+1}(c_j \mid \text{Pa}(c_j)).
\end{align*}
At each checkpoint interval $(t, t+1]$, the agent continuously updates these beliefs using newly collected evidence $\Delta \mathcal{D}^{t+1}$.
Since the exact computation of posteriors is generally infeasible for complex distributions, PAIR-Agent integrates free energy minimization with the Bayesian CFG and Markov blanket structure to infer approximate posterior beliefs. It utilizes the free energy principle (FEP)\cite{friston2010free} to align its internal model with observed system behavior continuously. At each checkpoint $t+1$, the agent computes an approximate posterior belief $Q^{t+1}(\mathbf{f})$ by minimizing the variational free energy, defined as 
$$F^{t+1} \triangleq \mathbb{E}_{Q^{t+1}(\mathbf{f})} \Big[\ln Q^{t+1}(\mathbf{f}) - \ln P^{t+1}(\mathbf{f}, \mathbf{x}^{t+1})\Big],$$
where $Q^{t+1}(\mathbf{f})$ approximates the posterior belief over the faults $\mathbf{f}=\{f_i\}_i$, given the contextual features $\mathbf{x}^{t+1}=\{\mathbf{x}_i^{t+1}\}_i$.
Minimizing $F^{t+1}$ updates these beliefs, refining the agent’s estimates under model uncertainty and capturing dependencies among faults.

To further refine fault-type attribution, PAIR-Agent partitions the inferred fault space into hardware-related and software-related components, 
\(\mathcal{V}_f^{t+1} = \mathcal{V}_{\text{HW}}^{t+1} \cup \mathcal{V}_{\text{SW}}^{t+1}\). 
The hardware set \(\mathcal{V}_{\text{HW}}^{t+1}\) captures physical and low-level operational indicators, such as temperature variations~\cite{Alfreds2025benchmarking}, power fluctuations, and connectivity degradation, while the software set \(\mathcal{V}_{\text{SW}}^{t+1}\) represents task failures, model divergence, resource contention, and execution-time anomalies. 
Using this decomposition, PAIR-Agent performs conditional marginalization to obtain $P^{t+1}(f_i \mid \mathcal{V}_{\text{HW}}^{t+1}) \quad \text{and} \quad P^{t+1}(f_i \mid \mathcal{V}_{\text{SW}}^{t+1}),$ which enables the agent to differentiate whether a detected fault most likely originates from hardware-level disturbances or software-driven inconsistencies. Importantly, this stage performs only \emph{fault inference}; the subsequent step handles all decision-making and recovery actions based on these inferred fault states.

\subsection{Active Inference for Healing}
Once fault inference identifies faults, PAIR-Agent employs active inference~\cite{sajid2021active} to autonomously plan and execute corrective actions. 
The agent evaluates a set of possible actions $a_t \in \mathcal{A}$ and selects the one that minimizes the expected free energy $G(\cdot)$: 
$$a_t^* = \argmin_{a_t} G(a_t).$$
The expected free energy for action $a_t$ is given by
\begin{align*}
    G(a_t) &\triangleq D_{\mathrm{KL}}[Q^{t+1}(\mathbf{x} \mid a_t) \parallel P^*(\mathbf{x})]
    \\&+\mathbb{E}_{Q^{t+1}(\mathbf{f} \mid a_t)}[ H[P(\mathbf{x} \mid \mathbf{f})]],
\end{align*}
where $P^*(\mathbf{x})$ encodes the preferred contextual features, $Q^{t+1}(\mathbf{f} \mid a_t)$ denotes the predicted posterior belief about the faults under action $a_t$, $P(\mathbf{x} \mid \mathbf{f})$ is the likelihood of observing the contextual features, given faults, and $Q^{t+1}(\mathbf{x} \mid a_t)\triangleq \sum_{\mathbf{f}}Q^{t+1}(\mathbf{f}\mid a_t)P(\mathbf{x}\mid \mathbf{f})$ is the predictive distribution over future contextual features under action $a_t$.
By minimizing \(G(a_t)\), PAIR-Agent selects actions that steer the system toward its preferred operational state while explicitly accounting for uncertainty in both hardware-level indicators (\(\mathcal{V}_{\text{HW}}^{t+1}\)) and software-level behavioral variables (\(\mathcal{V}_{\text{SW}}^{t+1}\)).
After executing the selected action, PAIR-Agent observes the resulting system state, updates posterior fault probabilities $P^{t+1}(f_i \mid \mathbf{x}_i^{t+1})$ with new evidence, and advances the checkpoint $\zeta^{t+1}$ to record the updated configuration and belief state. 
This closed-loop cycle of \emph{perception–inference–action–update} enables continual adaptation and resilience across analysis rounds.
During the healing process, PAIR-Agent not only redistributes tasks but also performs device-level recovery through automated reinitialization, firmware reloading, or adaptive configuration tuning. 
If such self-healing measures fail to restore stability, the agent isolates the faulty node and applies corrective actions at the continuum level $\mathcal{S}$, including workload redistribution across fog or cloud layers, rerouting through redundant network paths, managing dynamic load or thermal management, and escalation to human operators for physical interventions.

\section{Theoretical Results and Discussions}\label{sec4}
This section presents the theoretical foundations showing how PAIR-Agent achieves resilient operation in the DCC through principled probabilistic inference and cautious, uncertainty-aware action selection.
The results highlight four key properties: the locality and scalability of fault inference, the quality of the variational posterior approximation, the safety of the agent’s corrective actions, and the robustness of belief updates under missing or delayed logs.

\textbf{Result 1 (Locality and Scalability of Fault Inference).} For any fault variable $f_i$, PAIR-Agent only needs the variables in its Markov blanket $\mathcal{M}(f_i)$ (its parents, children, and co-parents) to update the belief $Q(f_i)$. Thus, the time and memory per fault update depend on $|\mathcal{M}(f_i)|$, and not on the total number of variables in the CFG. If the average blanket size is $B$, updating all $d$ faults in one round costs $O(d\cdot B)$, which stays manageable as the DCC grows, provided local neighborhoods stay small. This guarantees that the pipeline in Fig. \ref{fig1} scales due to inference being local by design.

\textbf{Result 2 (Best Possible Approximate Posterior).} At checkpoint $t+1$, PAIR-Agent selects $Q^{t+1}(\mathbf{f})$ by minimizing the variational free energy $F^{t+1}(Q)$. Since, $F^{t+1}(Q)$ differs from $D_{KL}[Q(\mathbf{f})\parallel P^{t+1}(\mathbf{f}\mid \mathbf{x}^{t+1})]$ only by a constant independent of $Q$ (i.e., $-\ln P^{t+1}(\mathbf{x}^{t+1})$), the update $Q^{t+1}=\argmin_{Q\in \mathcal{Q}}F^{t+1}(Q)$ is equivalently the closest distribution to the exact Bayesian posterior within the chosen variational family $\mathcal{Q}$. Therefore, given the current CFG and log data, PAIR-Agent computes the best posterior approximation it can represent, ensuring principled and internally consistent fault inference.

\textbf{Result 3 (Conservative Action Selection [Never Worse Than Doing Nothing]).} Assume that the action set includes a baseline "do nothing" action $a^\varnothing$, which keeps the current system configuration. Then, because PAIR-Agent selects $a_t^* = \argmin_{a_t} G(a_t)$, the chosen action never has higher expected free energy than $a^\varnothing$. Conversely, when an action is predicted to move $Q(\mathbf{x}\mid a)$ closer to preferences and/or reduce uncertainty, it will be selected. This guarantees that the agent will not take an action that it predicts will diverge the system from its operational objectives and not increase its model certainty.

\textbf{Result 4 (Robustness to Missing or Delayed Logs).} At checkpoint $t+1$, the belief update for any fault $f_i$ depends only on variables in its Markov blanket $\mathcal{M}(f_i)$. Therefore, missing or delayed logs for variables outside $\mathcal{M}(f_i)$ do not change the updated belief $Q^{t+1}(f_i)$. Since telemetry in DCC is often partial or out-of-order, this property makes PAIR-Agent's inference insensitive to unrelated data gaps and supports asynchronous, scalable updates: only logs inside the small local neighborhood $\mathcal{M}(f_i)$ can affect the diagnosis of $f_i$.

\section{Conclusions}\label{sec6}
This paper introduced the PAIR-Agent as a unified framework for achieving resilience in DCC systems. The PAIR-Agent agent performs: (i) constructing a CFG through device log analysis, (ii) identifying faults while managing uncertainties using Markov blankets and the FEP, and (iii) autonomously healing issues via active inference. Through its closed-loop cycle of observation, inference, and action, PAIR-Agent ensures adaptive stability, self-healing capability, and sustained operational continuity, marking a significant step toward autonomous and resilient DCC systems. Theoretical results confirm its scalability, principled and accurate fault inference, safe decision-making, and robustness to incomplete telemetry. Future work will extend this research with detailed and distributed implementation, empirical validation, and evaluation under a real-time testbed.

\bibliographystyle{IEEEtran}
\bibliography{ref.bib}

@ARTICLE{10562327,
  author={Fang, Honglin and Yu, Peng and Tan, Can and Zhang, Junye and Lin, Dahua and Zhang, Liyan and Zhang, Yong and Li, Wenjing and Meng, Luoming},
  journal={IEEE Network},
  title={Self-Healing in Knowledge-Driven Autonomous Networks: Context, Challenges, and Future Directions},
  year={2024},
  volume={38},
  number={6},
  pages={425-432},
  doi={10.1109/MNET.2024.3416850}}

@article{donta2024human_based,
	author = {Donta, Praveen Kumar and Sedlak, Boris and Murturi, Ilir and Casamayor-Pujol, V{\'\i}ctor and Dustdar, Schahram},
	doi = {10.1109/MIC.2024.3460908},
	journal = {IEEE Internet Computing},
	month = {09},
	title = {Human-based Distributed Intelligence in Computing Continuum Systems},
	year = {2025},
    volume = {29},
    number={2},
	doi = {10.1109/MIC.2024.3460908}
}

@article{sagor2024distressnet,
  title={{DistressNet-NG}: A resilient data storage and sharing framework for mobile edge computing in cyber-physical systems},
  author={Sagor, Mohammad and Haroon, Amran and Stoleru, Radu and Bhunia, Suman and Altaweel, Ala and Chao, Mengyuan and Jin, Liuyi and Maurice, Maxwell and Blalock, Roger},
  journal={ACM Transactions on Cyber-Physical Systems},
  volume={8},
  number={3},
  pages={1--31},
  year={2024},
  publisher={ACM New York, NY}
}

@article{donta2023governance,
  title={Governance and sustainability of distributed continuum systems: a big data approach},
  author={Donta, Praveen Kumar and Sedlak, Boris and Casamayor Pujol, Victor and Dustdar, Schahram},
  journal={Journal of Big Data},
  volume={10},
  number={1},
  pages={1--31},
  year={2023},doi={10.1186/s40537-023-00737-0},
  publisher={SpringerOpen}
}

@article{dustdar2022distributed,
  title={On distributed computing continuum systems},
  author={Dustdar, Schahram and Pujol, Victor Casamayor and Donta, Praveen Kumar},
  journal={IEEE Transactions on Knowledge and Data Engineering},
  volume={35},
  number={4},
  pages={4092--4105},
  year={2023},
doi={10.1109/TKDE.2022.3142856},
  publisher={IEEE}
}

@article{sajid2021active,
  title={Active inference: demystified and compared},
  author={Sajid, Noor and Ball, Philip J and Parr, Thomas and Friston, Karl J},
  journal={Neural computation},
  volume={33},
  number={3},
  pages={674--712},
  year={2021},
  publisher={MIT Press One Rogers Street, Cambridge, MA 02142-1209, USA journals-info~…}
}

@article{kirchhoff2018markov,
  title={The Markov blankets of life: autonomy, active inference and the free energy principle},
  author={Kirchhoff, Michael and Parr, Thomas and Palacios, Ensor and Friston, Karl and Kiverstein, Julian},
  journal={Journal of The royal society interface},
  volume={15},
  number={138},
  pages={20170792},
  year={2018},
  publisher={The Royal Society}
}

@article{svobodova2009resilient,
  title={Resilient distributed computing},
  author={Svobodova, Liba},
  journal={IEEE Transactions on Software Engineering},
  number={3},
  pages={257--268},
  year={2009},
  publisher={IEEE}
}

@article{friston2003dynamic,
  title={Dynamic causal modelling},
  author={Friston, Karl J and Harrison, Lee and Penny, Will},
  journal={Neuroimage},
  volume={19},
  number={4},
  pages={1273--1302},
  year={2003},
  publisher={Elsevier}
}

@article{cheikhrouhou2016secure,
  title={Secure group communication in wireless sensor networks: a survey},
  author={Cheikhrouhou, Omar},
  journal={Journal of Network and Computer Applications},
  volume={61},
  pages={115--132},
  year={2016},
  publisher={Elsevier}
}

@article{wang2025perturbation,
  title={Perturbation-resilient integer arithmetic using optical skyrmions},
  author={Wang, An Aloysius and Ma, Yifei and Zhang, Yunqi and Zhao, Zimo and Cai, Yuxi and Qiu, Xuke and Dong, Bowei and He, Chao},
  journal={Nature Photonics},
  pages={1--9},
  year={2025},
  publisher={Nature Publishing Group UK London}
}

@article{saleh2025agentic,
  title={Agentic {TinyML} for Intent-aware Handover in {6G} Wireless Networks},
  author={Saleh, Alaa and Morabito, Roberto and Tarkoma, Sasu and Lindgren, Anders and Pirttikangas, Susanna and Lovén, Lauri},
  journal={arXiv preprint arXiv:2508.09147},
  year={2025}
}

@article{scutari2018dirichlet,
  title={Dirichlet Bayesian network scores and the maximum relative entropy principle},
  author={Scutari, Marco},
  journal={Behaviormetrika},
  volume={45},
  number={2},
  pages={337--362},
  year={2018},
  publisher={Springer}
}

@INPROCEEDINGS{Alfreds2025benchmarking,
  author={Lapkovskis, Alfreds and Sedlak, Boris and Magnússon, Sindri and Dustdar, Schahram and Donta, Praveen Kumar},
  booktitle={2025 IEEE International Conference on Edge Computing and Communications (EDGE)}, 
  title={Benchmarking Dynamic SLO Compliance in Distributed Computing Continuum Systems}, 
  year={2025},
  volume={},
  number={},
  pages={93-102},
  doi={10.1109/EDGE67623.2025.00020}}

@article{friston2010free,
  title={The free-energy principle: a unified brain theory?},
  author={Friston, Karl},
  journal={Nature reviews neuroscience},
  volume={11},
  number={2},
  pages={127--138},
  year={2010},
  publisher={Nature publishing group}
}
\balance
\end{document}